# Portfolio Optimization with Entropic Value-at-Risk

**Amir Ahmadi-Javid[1] and Malihe Fallah-Tafti**

*Department of Industrial Engineering, Amirkabir University of Technology, Tehran, Iran*

**Abstract.** The *entropic value-at-risk* (EVaR) is a new coherent risk measure, which is an upper bound for both the *value-at-risk* (VaR) and *conditional value-at-risk* (CVaR). As important properties, the EVaR is strongly monotone over its domain and strictly monotone over a broad sub-domain including all continuous distributions, while well-known monotone risk measures, such as VaR and CVaR lack these properties. A key feature for a risk measure, besides its financial properties, is its applicability in large-scale sample-based portfolio optimization. If the negative return of an investment portfolio is a differentiable convex function, the portfolio optimization with the EVaR results in a differentiable convex program whose number of variables and constraints is independent of the sample size, which is not the case for the VaR and CVaR. This enables us to design an efficient algorithm using differentiable convex optimization. Our extensive numerical study shows the high efficiency of the algorithm in large scales, compared to the existing convex optimization software packages. The computational efficiency of the EVaR portfolio optimization approach is also compared with that of CVaR-based portfolio optimization. This comparison shows that the EVaR approach generally performs similarly, and it outperforms as the sample size increases. Moreover, the comparison of the portfolios obtained for a real case by the EVaR and CVaR approaches shows that the EVaR approach can find portfolios with better expectations and VaR values at high confidence levels.

## 1. Introduction

Portfolio optimization deals with forming a portfolio with maximum total return, or with minimum total loss (see Kolm et al. (2014) for a comprehensive review on practical portfolio optimization). There are different approaches to portfolio optimization, such as multi-objective optimization (Masmoudi & Abdelaziz, 2017), considering stochastic dominance constraints (see the review by Bruni et al. (2017)), accounting for ambiguity aversion (Källblad, 2017; Zhang et al., 2017), robustifying against uncertainties (Zymler et al., 2013; Wozabal, 2014; Doan et al., 2015), and incorporating investor's subjective views (Silva et al., 2017) or socially responsible investment criteria (Gasser et al., 2017).

A popular approach to portfolio optimization is the one broadly used in stochastic optimization (see, e.g., Ruszczyński and Shapiro (2006), Krokhmala et al. (2011), and Mastrogiacomo and Rosazza Gianin (2015)), where a single objective function representing the portfolio's risk is optimized subject to a set of deterministic constraints, which are reflecting the investor's policies; this approach can be extended to the multi-objective case when different stochastic objectives are scalarized using risk measures, see Gutjahr and Pichler (2016). In this approach, risk analysis concepts should be combined with optimization tools to obtain an optimal portfolio. A key factor in portfolio optimization using this approach is choosing a risk measure to scalarize the portfolio's risk. A risk measure is a real-valued functional that quantifies the degree of risk involved in a random outcome. Several papers have specified general properties for a suitable risk measure. The most important set of these properties might be given by Artzner et al. (1999) who axiomatically defined the family of *coherent* risk measures, which have four basic properties: translation invariance, monotonicity, subadditivity, and positive homogeneity.

[1] Corresponding author's email address: ahmadi_javid@aut.ac.ir

Unfortunately almost all popular risk measures introduced in the literature before 2000 lack some of the axiomatic properties required for coherency. For example, the classical mean-variance (or mean-standard-deviation) risk measure, inspired by Markowitz (1952), fails to be monotone, which make it economically meaningless. Another disadvantage of this risk measure is that it symmetrically penalizes profit and loss while risk is an asymmetric phenomenon. For this reason, the popularity of this risk measure dwindled and researchers showed increasing interest towards quantile-based measures, such as the *value-at-risk* (VaR). Although the VaR became so popular that it has been incorporated into industry regulations, it is not coherent and lacks the subadditivity property. This drawback entails both inconsistency with the well-accepted principle of diversification (diversification reduces risk) as well as computational intractability of optimization problems involving this risk measure. Working numerically with the VaR is unstable, and optimization models involving it are intractable in high dimensions (see, e.g., Sarykalin et al. (2008)).

To overcome these issues, recent literature on portfolio optimization has focused on coherent risk measures. Now there are several known coherent risk measures, such as spectral risk measures, but they cannot be used in large-scale portfolio optimization because their resulting optimization problems are not considered computationally tractable. Moreover, these risk measures may appear to have no clear interpretation for decision makers.

The *conditional value at risk* (CVaR) is a well-known coherent risk measure (Rockafellar & Uryasev, 2002; Rockafellar, 2007). Fortunately, linear portfolio optimization with the CVaR can be handled in reasonable times. Due to the simple *linear programming* (LP) reformulations for various CVaR-based linear portfolio optimization problems and suitable properties such as coherency, the CVaR rapidly gained in popularity (see, e.g., Mansini et al. (2014)). Nonetheless, the size of the resulting CVaR optimization problem by the LP reformulation is heavily dependent on the sample size, where choosing a large sample size leads to an LP model with a huge number of variables and constraints, which is computationally expensive. Ogryczak and Śliwiński (2011) showed that the computational efficiency of the simplex method can be dramatically improved when it is applied to the dual of the LP formulation of the CVaR optimization problem, but as the sample size increases this solution procedure similarly becomes very time consuming. To overcome this difficulty, some alternative solution methods have been recently explored to solve the original non-differentiable CVaR model, instead of the LP reformulation, using non-smooth optimization techniques. Lim et al. (2010) used techniques of non-differentiable optimization to effectively solve large-scale CVaR problem instances. Their numerical results show that using non-smooth optimization methods offers an advantageous computational alternative for the sample-based CVaR model with large sample sizes. There are many other papers that studied CVaR portfolio optimization problems, which are not detailed here (e.g., Beliakov and Bagirov (2006), Alexander et al. (2006), Tong et al. (2010), Künzi-Bay and Mayer (2006), Iyengar and Ma (2013), Espinoza and Moreno (2014), and Takano et al. (2015)) because the computational efficiency of their proposed methods cannot significantly improve the large-scale computational efficiency achieved in Lim et al. (2010) and Ogryczak and Śliwiński (2011). This conlsuion is only based on the sizes of the instances that these papers solved and reported, and can be carefully adjusted if a fair comparative numerical study is carried out, which remains as a future research area for the CVaR approach to linear portfolio optimization.

The *entropic value-at-risk* (EVaR) is a coherent risk measure that has been recently introduced in Ahmadi-Javid (2011, 2012a). Besides coherency, the EVaR has some other suitable properties. For example, contrary to other well-known monotone risk measures, such as the VaR and the CVaR, the EVaR is strongly and strictly monotone (see Section 2). Extensions of the EVaR, and g-entropic risk measures introduced together with the EVaR in Ahmadi-Javid (2011, 2012a), have been studied in Ahmadi-Javid (2012c), and Breuer and Csiszár (2013, 2016). New features of the EVaR are investigated in Pichler (2017) and Delbaen (2018). Norms and Banach spaces induced by the EVaR are studied by Ahmadi-Javid and Pichler (2017). The EVaR is applied in different fields, such as machine learning (Ahmadi-Javid, 2012d), learning-based autonomous systems (Axelrod et al., 2016), approximate statistical models (Watson & Holmes, 2016), and distributionally robust optimization (Postek et al., 2016).

When using the model-based approach to represent input data, portfolio optimization with the EVaR can be done efficiently for a broad range of return rates with known distributional properties; for example, when return rates follow arbitrary independent distributions, or when they follow a generalized linear multi-factor model, inspired by Sharp (1963). Note that these cases cannot be handled for the VaR or CVaR (for more details see Section 4 in Ahmad-Javid (2012a)).

When using historical or Monte Carlo simulation approaches to represent input data, in portfolio optimization, it is desirable to include as many samples as possible so that the problem reliably reflects the underlying distribution of returns. This paper for the first time examines the computational efficiency of sample-based portfolio optimization with the EVaR in large scales. Fortunately, dissimilar to the CVaR case, when the portfolio return is linear or differentiable convex, the problem with the EVaR can be formulated as a differentiable convex optimization problem where the total number of variables and constraints is independent of the sample size. Thus, one can solve the resulting optimization problem very reliably and efficiently using various methods developed for differentiable convex programming. In addition to the fact that the EVaR has appealing features such as coherency and more desirable monotonicity properties, our main goal is to show that sample-based portfolio optimization with this risk measure is practical and can be done very efficiently.

The remainder of this paper is organized as follows: Section 2 reviews preliminaries on risk measures and the EVaR. Section 3 introduces the sample-based portfolio optimization model with a given risk measure, which is detailed for the CVaR and the EVaR. Section 4 presents a primal-dual interior-point algorithm to solve the sample-based portfolio optimization with the EVaR. Section 5 reports a comprehensive computational study to assess the efficiency of the proposed algorithm. This section also compares computational times required for the sample-based portfolio optimization with the CVaR and the EVaR. Finally, Section 6 closes the paper with conclusions.

## 2. Risk measures and properties of EVaR

A risk measure is a function $\rho$ which assigns a real value to a random variable $X$ in order to quantify its risk degree. To precisely define the concept of a risk measure, consider a probability space $(\Omega, \mathbf{F}, P)$ where $\Omega$ is a set of all simple events, $\mathbf{F}$ is a $\sigma$-algebra of subsets of $\Omega$, and $P$ is a probability measure on $\mathbf{F}$. Also, let $\mathbf{L}_0$ be the set of all Borel measurable functions (random variables) $X: \Omega \to \mathbb{R}$, and $\mathbf{X} \subseteq \mathbf{L}_0$ be the model space, which is a subspace including all real numbers (constant functions). Then, a risk measure is defined as $\rho: \mathbf{X} \to \overline{\mathbb{R}}$, where $\overline{\mathbb{R}} = \mathbb{R} \cup \{-\infty, +\infty\}$ is the extended real line. The literature introduces several properties for a suitable risk measure. The following may be the most important properties for the risk measure $\rho$:

(P1) Translation Invariance: $\rho(X + c) = \rho(X) + c$ for any $X \in \mathbf{X}$ and $c \in \mathbb{R}$
(P2) Subadditivity: $\rho(X_1 + X_2) \leq \rho(X_1) + \rho(X_2)$ for all $X_1, X_2 \in \mathbf{X}$
(P3) Monotonicity: If $X_1, X_2 \in \mathbf{X}$ and $X_1 \leq X_2$, then $\rho(X_1) \leq \rho(X_2)$
(P4) Positive Homogeneity: $\rho(\lambda X) = \lambda \rho(X)$ for all $X \in \mathbf{X}$ and $\lambda \geq 0$.

A risk measure is called *coherent*, if it satisfies the properties P1–P4 (Artzner et al., 1999). Our definition of a coherent risk measure is taken from the operations research literature (Ruszczyński and Shapiro, 2006), where $X \in \mathbf{X}$ is viewed as a loss or cost. However, in the finance literature, $X$ often represents a gain or profit, and thus a risk measure $\rho$ is called coherent if the functional $\psi(X) = \rho(-X)$ satisfies the above four properties.

Now we give the definitions of some of the most popular risk measures that have been used by researchers. For $p \geq 1$, let $\mathbf{L}_p \subseteq \mathbf{L}_0$ be the set of all random variables $X$ for which $\mathrm{E}(|X|^p) = \int |X|^p dP < \infty$, let $\mathbf{L}_\infty \subseteq \mathbf{L}_0$ denote the set of all bounded random variables, and let $\mathbf{L}_M$ stand for the set of all random variables $X$ whose moment-generating function $M_X(\theta) = \mathrm{E}(e^{\theta X})$ is finite for all $\theta \in \mathbb{R}$. Note that $\mathbf{L}_\infty \subseteq \mathbf{L}_M \subseteq \mathbf{L}_p \subseteq \mathbf{L}$ for all $p \geq 1$. The expectation and worst-case risk measures are two elementary coherent risk measures defined by

$$\rho_\mathrm{E}(X) := \mathrm{E}(X), \quad X \in \mathbf{L}_1$$
$$\rho_\mathrm{W}(X) := \mathrm{ess\,sup}(X), \quad X \in \mathbf{L}_\infty$$

where $\mathrm{ess\,sup}(X) = \inf\{t \in \mathbb{R}: \Pr(X \leq t) = 1\}$ denotes the essential supremum of $X$. Unfortunately, most other well-known risk measures that have been used in the literature for decades are not coherent. For example, the mean-standard-deviation risk measure

$$\mathrm{MV}_\lambda(X) := \mathrm{E}(X) + \lambda \cdot \mathrm{SD}(X), \quad X \in \mathbf{L}_2, \quad \lambda > 0$$

is not generally monotone. The subadditivity property does not hold for the *value-at-risk* (VaR), where VaR with confidence level $1 - \alpha$ (or at risk level $\alpha$) is given by

$$\mathrm{VaR}_{1-\alpha}(X) := \inf\{t \in \mathbb{R}: \Pr(X \leq t) \geq 1 - \alpha\}, \quad X \in \mathbf{L}_0, \quad \alpha \in [0,1].$$

The *conditional value-at-risk* (CVaR, also called *average value-at-risk* or *expected shortfall*), is an important coherent risk measure that has gained popularity in recent years. Actually it is the smallest law-invariant coherent risk measure that is greater than the VaR (see Remark 3 below). The CVaR with confidence level $1-\alpha$ (or at risk level $\alpha$) is defined as follows:

$$\text{CVaR}_{1-\alpha}(X) := \inf_{t \in \mathbb{R}}(t + \frac{1}{\alpha}\text{E}[X-t]_+), \quad X \in \mathbf{L}_1, \alpha \in (0,1]$$

where $[s]_+ = \max\{0, s\}$. This measure can be represented by the VaR as follows:

$$\text{CVaR}_{1-\alpha}(X) = \frac{1}{\alpha}\int_0^\alpha \text{VaR}_{1-t}(X)dt.$$

This roughly means that $\text{CVaR}_{1-\alpha}(X)$ is the average of the worst $\alpha\%$ of values of $X$. For a detailed comparison between the VaR and the CVaR in risk management and optimization, see Sarykalin et al. (2008).

The *entropic value-at-risk* (EVaR) is a new coherent risk measure recently introduced and studied by Ahmadi-Javid (2011, 2012a). The EVaR with confidence level $1-\alpha$ (or at risk level $\alpha$) is defined as

$$\text{EVaR}_{1-\alpha}(X) := \inf_{\theta>0}\{\theta^{-1}\ln(M_X(\theta)/\alpha)\}, \quad X \in \mathbf{L}_M, \alpha \in (0,1]. \tag{2-1}$$

For a comparison, the VaR, CVaR, and EVaR for a normal distribution, $X \sim N(\mu, \sigma^2)$ are as follows:

$$\text{VaR}_{1-\alpha}(X) = \mu + z_\alpha \sigma$$

$$\text{CVaR}_{1-\alpha}(X) = \mu + \frac{\phi(z_\alpha)}{\alpha}\sigma$$

$$\text{EVaR}_{1-\alpha}(X) = \mu + \sqrt{-2\ln\alpha}\,\sigma$$

where $\phi(.)$ and $z_\alpha$ are the density function and the upper $\alpha$-percentile of the standard normal distribution, respectively. We see that under the normality assumption, all of these risk measures can be expressed in terms of the mean and variance, and they are actually equivalent to the mean-standard-deviation risk measure for different values of $\lambda$.

**Remark 1.** As considered in Ahmadi-Javid and Pichler (2017), the EVaR is well-defined on a larger vector space denoted by $\boldsymbol{E}^{**}$, compared to $\mathbf{L}_M$, which is called the *bidual entropic space* and defined as the set of random variables $X$ for which the moment-generating function $M_{|X|}(\theta) = \text{E}(e^{\theta|X|})$ exists for some $\theta > 0$. In Ahmadi-Javid and Pichler (2017), $\mathbf{L}_M$ is denoted by $\boldsymbol{E}$, and it is shown that $\boldsymbol{E}^{**}$ is the bidual space of $\boldsymbol{E}$ with respect to the norm induced by the EVaR. This paper also provides deep analysis of norms and Banach spaces related to the EVaR, and their connections with the theory of Orlicz spaces. Note that the natural domain of the EVaR is a convex cone, not a vector space. ∎

The EVaR is the tightest upper bound that one can find using the Chernoff inequality for the VaR (and CVaR). Its dual and Kusuoka representations are given based on the relative entropy, also known as Kullback-Leibler divergence, which unseals why this risk measure was called Entropic VaR (see Ahmadi-Javid (2012a,b), Delbaen (2018), and Ahmadi-Javid and Pichler (2017)).

Optimization with the EVaR is computationally tractable for a wide class of quantitative risks, which are not efficiently computable for the VaR and CVaR (Ahmadi-Javid, 2012a). Moreover, this risk measure has the important property of strong monotonicity, defined below, which does not hold for other popular (coherent or non-coherent) monotone risk measures such as VaR or CVaR. Recall that the mean-variance and mean-standard-deviation are instances of risk measures that became popular after the seminal work of Markowitz (1952), but they have been overlooked by many researchers because of their lack of monotonicity.

A risk measure $\rho$ is called *strongly monotone* if it holds that $\rho(X) < \rho(Y)$ for any pair of random variables $X$ and $Y$ in the domain of $\rho$ that satisfy the conditions

C1) $X \geq Y$

C2) $\Pr\{X > Y\} > 0$

C3) $\operatorname{esssup} X > \operatorname{esssup} Y$ or $\operatorname{esssup} X = \operatorname{esssup} Y = +\infty$.

In fact, these conditions imply the strict preference of $X$ over $Y$ by the risk measure $\rho$. Apparently, strong monotonicity is a desirable property for a risk measure. The next theorem proves that the EVaR has this property.

**Theorem 1. (Strong monotonicity of the EVaR)** Let $X$ and $Y$ be random variables in $\mathbf{L}_M$ that satisfy C1, C2, and C3. Then, $\text{EVaR}_{1-\alpha}(X) > \text{EVaR}_{1-\alpha}(Y)$ for any $\alpha \in (0,1]$.

**Proof.** Defining $f_X(t, \alpha) = t \log \text{E}(\exp\{t^{-1}X\}) - t \ln \alpha$, the EVaR can be rewritten as

$$\text{EVaR}_{1-\alpha}(X) = \inf_{t>0}\{f_X(t, \alpha)\}.$$

Conditions C1 and C2 imply $\text{E}(\exp\{t^{-1}X\}) > \text{E}(\exp\{t^{-1}Y\})$ for $t > 0$, then for $t > 0$

$$f_X(t, \alpha) = t \log \text{E}(\exp\{t^{-1}X\}) - t \ln \alpha > f_Y(t, \alpha) = t \log \text{E}(\exp\{t^{-1}Y\}) - t \ln \alpha.$$

Moreover, by C3

$$\lim_{t \to 0} f_X(t, \alpha) = \operatorname{esssup} X > \lim_{t \to 0} f_Y(t, \alpha) = \operatorname{esssup} Y, \text{ or}$$

$$\lim_{t \to 0} f_X(t, \alpha) = \lim_{t \to 0} f_Y(t, \alpha) = +\infty.$$

Thus, $\text{EVaR}_{1-\alpha}(X) = \inf_{t>0}\{f_X(t, \alpha)\} > \text{EVaR}_{1-\alpha}(Y) = \inf_{t>0}\{f_Y(t, \alpha)\}$. This ends the proof. ∎

**Remark 2. (VaR and CVaR are not strongly monotone)** Though both VaR and CVaR are monotone, they lack the strong monotonicity property. To illustrate this, let us define the following random losses:

$$X \sim N(\mu, \sigma^2), Y_M = \begin{cases} X & X \geq \text{VaR}_{1-\alpha}(X) \\ X - M & X < \text{VaR}_{1-\alpha}(X) \end{cases}$$

where $M$ is an arbitrary positive number ($X$ can have any other continuous distribution with finite mean). These random variables satisfy conditions C1, C2, and C3, so $Y_M$ strictly dominates $X$ for any $M > 0$. $Y_M$ obviously outperforms $X$ for any $M > 0$, and $Y_M$ is preferred over $Y_{M'}$ for any $M' < M$. However, both the VaR and CVaR return the same values for these random variables, i.e.,

$$\text{VaR}_{1-\alpha}(X) = \text{VaR}_{1-\alpha}(Y_M) = \mu + z_\alpha \sigma$$

$$\text{CVaR}_{1-\alpha}(X) = \text{CVaR}_{1-\alpha}(Y_M) = \mu + \frac{\phi(z_\alpha)}{\alpha}\sigma$$

for any small or large positive number $M$. ∎

The reason behind that the CVaR is not strongly monotone is that this risk measure only considers the losses beyond the VaR and does not have any control on the losses less than the VaR. However, the EVaR incorporates the losses less and greater than the VaR simultaneously. It is why the EVaR is always larger than the CVaR at the same confidence level.

**Remark 3. (No VaR-bounded risk measure is strictly monotone)** A risk measure $\rho$ is called *strictly monotone* if for any random variables $X$ and $Y$ in the domain of $\rho$ that satisfies conditions C1 and C2, given before stating Theorem 1, it holds that $\rho(X) < \rho(Y)$ (see Section 4.2 in Delbaen (2000)). Moreover, a risk measure $\rho$ is called *VaR-bounded* if $\text{VaR}_{1-\alpha}(X) \leq \rho(X)$ for all $X$ in the domain of $\rho$. The smallest law-invariant VaR-bounded coherent risk measure is the CVaR (see Section 6 in Delbaen (2002)), where a risk measure $\rho(X)$ is called law-invariant if the value of $\rho(X)$ only depends on the distribution of $X$. The EVaR is also a law-invariant VaR-bounded risk measure.

It can be clearly seen that there is no strictly monotone VaR-bounded risk measure $\rho$. Indeed, for any non-constant random variable $Y$ and risk level $\alpha$ with $\text{VaR}_{1-\alpha}(Y) = esssup\,(Y)$, every VaR-bounded risk measure $\rho$ becomes equal to $esssup\,(X)$, and thus the risk measure $\rho$ losses the strict monotonicity property for the pair $X = esssup\,(Y)$ and $Y$. Therefore, none of the CVaR and EVaR, or any other VaR-bounded monotone risk measures, is strictly monotone. ∎

**Remark 4. (EVaR is strictly monotone over a broad sub-domain)** It can be shown that the EVaR is strictly monotone over the set of random variables with a distribution that takes its essential supremum with probability less than $\alpha$. This immediately implies that the EVaR is strictly monotone over the set of random variables with continuous distributions, while both VaR and CVaR are not (see the example given in Remark 2). ∎

**Remark 5. (Appropriate strictly-monotone coherent risk measures)** One can easily construct coherent risk measures that are strictly monotone. Examples are spectral risk measures whose distortion functions are strictly positive, or risk measures which are the sum of a monotone risk measure and the expectation (see Propositions 6.37 and 6.38 in Shapiro et al. (2014)). However, there are two major problems with such risk measures. The first is that the resulting risk measures may lack a clear interpretation for decision makers. The second is that, unlike the CVaR or the EVaR, they may not be used in large-scale optimization problems. An important open question in the modern risk measure theory would be to define a coherent risk measure with an acceptable interpretation and strict monotonicity property over its domain, such that it can be efficiently used in stochastic optimization. The EVaR is a partial answer to this open question, as it is strictly monotone over a broad sub-domain, including random variables with continuous random variables (and strongly monotone over its domain). ∎

## 3. Portfolio Optimization with EVaR

Let $w \in \mathbb{R}^n$ denote a portfolio vector indicating the fraction of investment of some available budget (say 1) in each one of $n$ financial instruments. Let $R$ be a $k$-dimensional real-valued random vector with a known probability distribution, representing $k$ risk factors required to specify the portfolio return. The random vector $R$ is modeled over the underlying probability space $(\Omega, \mathbf{F}, P)$. The loss of the portfolio (the negative return) is also a random variable denoted by $G(w, R)$, which depends on the vector of risk factors $R$ and portfolio vector $w$. Assuming that no short positions are allowed, a sensible portfolio can be determined by solving the following optimization problem:

$$\begin{aligned} \min \,& \rho(G(w,R)) \\ s.t. \,& w^T \mathbf{1} = 1 \\ & w \geq 0 \\ & w \in D \end{aligned} \quad (3\text{-}1)$$

where $\rho$ is a risk measure, representing the risk of the portfolio $w$, and $D \subset \mathbb{R}^n$ is a compact set, representing user-specified requirements, such as having a portfolio with the mean return at least $r_{min}$, i.e., $D = \{w \in \mathbb{R}^n : \mathrm{E}(-G(w,R)) \geq r_{min}\}$. Note that linear transaction costs can be also added to the above setting using $O(n)$ number of variables and linear constraints. Hence, the following analysis remains correct if transaction costs are also included in the model.

To have a well-defined problem, for all $w \in D$, it is required to assume that $G(w,.)$ is a Borel measurable function and that the objective function in (3-1) is finite; the random variable $G(w,R)$ always lies in the model space considered for the risk measure $\rho$.

It can be proven that the objective function of the problem (3-1) is convex if the risk measure $\rho$ is coherent and $G(.,s)$ is convex for all $s \in Supp_R$, where $Supp_R$ denotes the support of the random vector $R$.

**Theorem 2**. Let $\rho$ be a coherent risk measure, $D$ be a convex set, and $G(.,s)$ be convex for all $s \in Supp_R$. Then, the problem (3-7) is a convex optimization problem.

**Proof**. To accept the claim, the convexity of the objective function $h(w) = \rho(G(w,R))$ must be shown. For any pair of points $w_1$ and $w_2$ that are feasible to the constraints of (3-1), and every $\tau \in (0,1)$, one has

$$\begin{aligned} & h(\tau w_1 + (1-\tau)w_2) \\ &= \rho\big(G(\tau w_1 + (1-\tau)w_2, R)\big) \leq \rho\big(\tau G(w_1, R) + (1-\tau) G(w_2, R)\big) \\ &\leq \rho\big(\tau G(w, R)\big) + \rho\big((1-\tau) G(w_2, R)\big) \\ &= \tau \rho\big(G(w_1, R)\big) + (1-\tau) \rho\big(G(w_2, R)\big) = \tau h(w_1) + (1-\tau) h(w_2). \end{aligned}$$

The first inequality follows from the monotonicity of the risk measure $\rho$ and the convexity of the function $G$. The second follows from the subadditivity of the risk measure $\rho$. Finally, the last equality holds by the positive homogeneity of risk measure $\rho$. This completes the proof. ∎

It what follows, to simplify our discussion, unless otherwise stated, it is assumed that the set $D$ can be expressed by a polynomial-size number of differentiable convex constraints, i.e.,

$$D = \{w \in \mathbb{R}^n : h_p(w) \leq 0, p = 1, \ldots, m\}$$

where $h_p(.), p = 1, \ldots, m$ are differentiable convex functions and where $m$ grows polynomially with the number of instruments $n$. Moreover, in the sequel, the type of differentiability is assumed to be the same for all of the functions under consideration. This type will be further specified whenever it is required.

In practice, the collected (empirical or synthesized) samples are commonly used to express the joint probability distribution of risk factors $R$. We denote the samples from the random vector $R$ by $a^j = (a_1^j, \ldots, a_k^j)^T$, with probability $p_j$, $j = 1, \ldots, N$, where $N$ is the *sample size*. This setting is here called *sample-based setting* and will be considered in the rest of the paper. Note that in most practical cases, all probabilities are set to $N^{-1}$.

## 3.1. Problem (3-1) for CVaR under sample-based setting

When $\rho$ is set to the CVaR, under the sample-based setting, the problem (3-1) can be rewritten as the following non-differentiable convex optimization problem:

$$\min t + \alpha^{-1} \sum_{j=1}^{N} p_j \max\{G(w, a^j) - t, 0\}$$
$$s.t. \ w^T 1 = 1 \qquad \qquad (3\text{-}2)$$
$$w \in D$$
$$w \geq 0, t \in \mathbb{R}.$$

A possible way to obtain a differentiable formulation is to introduce auxiliary variables $z_j$, $j = 1, \ldots, N$, as follows:

$$\min t + \alpha^{-1} \sum_{j=1}^{N} p_j z_j$$
$$s.t. \ z_j \geq G(w, a^j) - t, \quad j = 1, \ldots, N \qquad (3\text{-}3)$$
$$w^T 1 = 1$$
$$w \in D$$
$$w \geq 0, z \geq 0, t \in \mathbb{R}.$$

The number of constraints and variables of this program depends on the sample size $N$. This shows, even if the functions $G(w, a^j), j = 1, \ldots, N$, are differentiable and convex, this formulation cannot be efficiently solvable for a large sample size. However, for the special case of linear portfolio optimization with the CVaR, various algorithms have been developed, which are briefly discussed below.

For the linear case $G(w, R) = -R^T w$, where $R$ is now the vector of the return rates of the instruments, with $D = \mathbb{R}_+^n$, the formulation (3-3) is given by

$$\min t + \alpha^{-1} \sum_{j=1}^{N} p_j z_j$$
$$s.t. \ z_j \geq -(a^j)^T w - t, \quad j = 1, \dots, N$$
$$w^T 1 = 1$$
$$w \geq 0, z \geq 0, t \in \mathbb{R}.$$

(3-4)

Solving (3-4) for large sample size $N$ is very time consuming in practice. An alternative method to more efficiently solve the above problem is to solve its dual program (Ogryczak and Śliwiński, 2011), given by

$$\max \xi$$
$$s.t. \ \xi + \sum_{j=1}^{N} a_i^j u_j \leq 0, \quad i = 1, \dots, n$$
$$0 \leq u_j \leq \alpha^{-1} p_j, \quad j = 1, \dots, N$$
$$\xi \in \mathbb{R}.$$

(3-5)

Similarly, solving the dual (3-5) is not practically possible for very large sample sizes. Thus, non-smooth optimization algorithms can be used based on the following non-differentiable formulation:

$$\min t + \alpha^{-1} \sum_{j=1}^{N} p_j \max\{-(a^j)^T w - t, 0\}$$
$$s.t. \ w^T 1 = 1$$
$$w \geq 0, t \in \mathbb{R}.$$

(3-6)

See Lim et al. (2010) and references therein for more details on such algorithms.

### 3.2. Problem (3-1) for EVaR under sample-based setting

If risk measure $\rho$ in the problem (3-1) is replaced by the EVaR, under the sample-based setting, the following optimization problem is obtained:

$$\min \ t \ln\left(\sum_{j=1}^{N} p_j e^{t^{-1} G(w, a^j)}\right) - t \ln \alpha$$
$$s.t. \ w^T 1 = 1$$
$$w \in D$$
$$w \geq 0, t > 0,$$

(3-7)

where variable $\theta$ in the definition of the EVaR is replaced with $t^{-1}$ to convexify the resulting problem. The next theorem shows that this problem is a convex program if the negative return is convex in decision variables $w$.

**Theorem 3**. If the functions $G(w, a^j), j = 1, ..., N,$ are (differentiable) convex in $w$, the problem (3-7) is a (differentiable) convex program.

**Proof**. It is sufficient to prove that the objective function is convex. Consider the following function:

$$h_1(v) = \ln\left(\sum_{j=1}^{N} p_j e^{v_j}\right) - \ln \alpha$$

which is a convex function in $v$. The perspective of this function, given by $h_2(v, t) = t h_1(t^{-1}v)$, is jointly convex in $(v, t)$. Because $h_2(v, t)$ is an increasing function in $v$ and the functions $G(w, a^i)$, $i = 1, ..., N$, are convex functions, the composite function $h_3(w, t) = h_2(G(w, a^1), ..., G(w, a^N), t)$ is also a convex function in $(w, t)$. This completes the proof as $h_3(w, t)$ is the objective function of (3-7). The proof for the differentiability is straightforward. ∎

**Remark 6. (Differentiability of EVaR-based portfolio optimization problems)** One should note that if the functions $G(w, a^j), j = 1, ..., N$, are differentiable (convex), the resulting problem (3-7) is a differentiable (convex) program whose number of variables and constraints <u>is independent of the sample size $N$</u>. However, this does not hold for the CVaR case, which can be seen from (3-3) for the linear case, where $N$ variables and $2N$ constraints are added to convert the non-differentiable formulation (3-2) to a differentiable program. It is an important feature for the EVaR because the most powerful and stable convex optimization algorithms are those developed for differentiable convex programs. ∎

## 4. Solution algorithm

Theorem 2 indicates that the problem (3-7) becomes a differentiable convex program when $G(w, a)$ is a differentiable convex function for all samples $a \in Supp_R$. This means that one can use existing differentiable convex optimization methods for solving this problem. Differentiable convex optimization problems, beyond having rich theoretical advantages, can be solved very reliably and efficiently using interior-point methods. The primal-dual methods are among the most efficient interior-point methods developed for solving differentiable convex optimization problems (Boyd and Vandenberghe, 2004).

This section shows how the primal-dual method can be successfully used to solve very large-sized instances of the problem (3-7). To apply the primal-dual method, the functions $G(w, a^j)$, $j = 1, ..., N$, and $h_p(.), p = 1, ..., m$ must be continuously twice differentiable convex function, and $D$ must include an interior point. Other first-order methods, such as the gradient method and its extensions can be used if some of the functions $G(w, a^j)$, $j = 1, ..., N$, and $h_p(.), p = 1, ..., m$ are only once differentiable.

Moreover, because an EVaR-based portfolio optimization problem is differentiable and convex with only <u>one additional variable</u> $t$, provided that the functions $G(w, a^j)$, $j = 1, ..., N$, and $h_p(.), p = 1, ..., m$ are differentiable and convex, one can easily use the existing general-purpose optimization packages (e.g., GAMS that uses many nonlinear optimization solvers such as CONOPT) to efficiently solve moderate-sized instances of the problem, instead of programming a specialized algorithm. However, to achieve more computational efficiency in large scales, it is common to use specialized algorithms, as the literature indicates that large-sized instances of the CVaR-based linear portfolio optimization can be efficiently solved using specialized algorithms.

One should carefully note that our aim is not to design the best possible algorithm to solve our problem because this paper for the first time provides a specialized algorithm for this problem. Our main aim is to show that, using a differentiable optimization method, the portfolio optimization with the EVaR can be efficiently applied in large scales despite the nonlinearity of the resulting model, where the computational efficiency is compared with that of general-purpose optimization solvers and portfolio optimization with the CVaR. Future studies could be conducted to develop more efficient algorithms based on other existing methods or under different assumptions on the structural properties of the return function and constraints (as it is done for the CVaR by many papers, see the introduction for a short review).

Section 4.1 briefly provides an introduction to the primal-dual method, and Section 4.2 shows how this method can be used to design an algorithm for the portfolio optimization with the EVaR.

### 4.1. Primal-dual interior-point method

One of the theoretical advantages of a convex optimization problem is that for the associated dual problem under a mild condition (for example, the Slater's constraint qualification) the duality gap is zero, which implies that the primal and dual problems have the same optimal values. This leads to efficient primal-dual interior point methods that iteratively and simultaneously update both the primal and dual variables. In this subsection, we briefly describe a standard primal-dual interior-point method (for more details see Boyd and Vandenberghe (2004)).

Consider the following standard form of a convex optimization problem:
$$\min f_0(\boldsymbol{x})$$
$$s.t.\ f_i(\boldsymbol{x}) \leq 0, \quad i = 1, \dots, m \qquad (4\text{-}1)$$
$$\boldsymbol{Gx} = \boldsymbol{h},$$

where $f_0, \dots, f_m: \mathbb{R}^n \to \mathbb{R}$ are convex and continuously twice differentiable, and $\boldsymbol{G} \in \mathbb{R}^{p \times n}$ with rank($\boldsymbol{G}$) = $p < n$. Assume that $\boldsymbol{\lambda} \in R^m$ and $\boldsymbol{\nu} \in R^m$ are dual variables related to the inequality and equality constraints of the above problem, respectively, and define $\boldsymbol{y} = (\boldsymbol{x}, \boldsymbol{\lambda}, \boldsymbol{\nu})$. For solving this problem, the standard primal-dual interior-point method produces a minimizing sequence $\boldsymbol{y}^{(k)}, k = 1,2,\dots$, where $\boldsymbol{y}^{(k+1)} = \boldsymbol{y}^{(k)} + \gamma^{(k)} \Delta \boldsymbol{y}^{(k)}$ with the search direction $\Delta \boldsymbol{y}^{(k)} \in \mathbb{R}^n$ and step size $\gamma^{(k)} \in \mathbb{R}_+$. Focusing on a typical iteration of the algorithm, we use the lighter notation $\boldsymbol{y} := \boldsymbol{y} + \gamma \Delta \boldsymbol{y}$ in place of the above ones.

#### 4.1.1. Primal-dual search direction

Consider the modified KKT conditions $r_z(\boldsymbol{x}, \boldsymbol{\lambda}, \boldsymbol{\nu}) = 0$, where $r_z(\boldsymbol{x}, \boldsymbol{\lambda}, \boldsymbol{\nu})$ is defined as

$$r_z(\boldsymbol{x}, \boldsymbol{\lambda}, \boldsymbol{\nu}) = \begin{bmatrix} \nabla f_0(\boldsymbol{x}) + Df(\boldsymbol{x})^T \boldsymbol{\lambda} + \boldsymbol{G}^T \boldsymbol{\nu} \\ -\text{diag}(\boldsymbol{\lambda}) f(\boldsymbol{x}) - (1/z)\mathbf{1} \\ \boldsymbol{Gx} - \boldsymbol{h} \end{bmatrix}, \qquad (4\text{-}2)$$

with $z > 0$. Here $f: \mathbb{R}^n \to \mathbb{R}^m$ and its derivative matrix $Df$ are given by

$$f(x) = \begin{bmatrix} f_1(x) \\ \vdots \\ f_m(x) \end{bmatrix}, Df(x) = \begin{bmatrix} \nabla f_1(x)^T \\ \vdots \\ \nabla f_m(x)^T \end{bmatrix}. \tag{4-3}$$

If $x, \lambda, \nu$ satisfy $r_z(x, \lambda, \nu) = 0$, then $x$ is primal feasible, and $\lambda, \nu$ are dual feasible, with duality gap $m/z$. The first block component of $r_z$, $r_{\text{dual}} = \nabla f_0(x) + Df(x)^T \lambda + G^T \nu$, is called the dual residual, the last block component, $r_{\text{pri}} = Gx - h$, is called the primal residual, and the middle block, $r_{\text{cent}} = -\text{diag}(\lambda)f(x) - (1/z)\mathbf{1}$, is called the centrality residual.

The search direction in the primal-dual interior-point method, denoted by $\Delta y_{\text{pd}} = (\Delta x_{\text{pd}}, \Delta \lambda_{\text{pd}}, \Delta \nu_{\text{pd}})$, is defined as the solution of the system

$$\begin{bmatrix} \nabla^2 f_0(x) + \sum_{i=1}^{m} \lambda_i \nabla^2 f_i(x) & Df(x)^T & G^T \\ -\text{diag}(\lambda)Df(x) & -\text{diag}(f(x)) & 0 \\ G & 0 & 0 \end{bmatrix} \begin{bmatrix} \Delta x \\ \Delta \lambda \\ \Delta \nu \end{bmatrix} = - \begin{bmatrix} r_{\text{dual}} \\ r_{\text{cent}} \\ r_{\text{pri}} \end{bmatrix}, \tag{4-4}$$

which is obtained from Newton's method that is applied to the modified KKT equations $r_z(x, \lambda, \nu) = 0$, with $r_z$ defined in (4-2).

### 4.1.2. The surrogate duality gap

In the standard primal-dual interior-point method, as the iterates $x^{(k)}$, $\lambda^{(k)}$, and $\nu^{(k)}$ are not necessarily feasible, one cannot easily evaluate a duality gap $\eta^{(k)}$ associated with step $k$. Instead, the surrogate duality gap, for any $x$ that satisfies $f(x) < 0$ and $\lambda \geq 0$, is defined as

$$\hat{\eta}(x, \lambda) = -f(x)^T \lambda. \tag{4-5}$$

The surrogate gap $\hat{\eta}$ is the duality gap, if $x$ is primal feasible and $\lambda, \nu$ are dual feasible; or, equivalently, the primal and dual residuals are equal to zero.

### 4.1.3. A primal-dual interior-point algorithm

This subsection now describes a primal-dual interior-point algorithm, which is called the *PD algorithm* here, based on the standard primal-dual method explained above. More sophisticated algorithms of this type differ from the PD algorithm in the strategy of selecting $z$ (which is crucial to have super-linear asymptotic convergence) and the line search. We refer to the survey of Forsgren et al. (2002) for details and references of different variations. For a set of complex constraints, this algorithm needs a suitable starting point. The calculation of this point is discussed in Section 4.1.3.2.

**The PD algorithm:**
*Input* $x$ that satisfies $f_1(x) < 0, \ldots, f_m(x) < 0$, $\lambda > \mathbf{0}$, any $\nu$, $\mu > 1$, $\epsilon_{feas} > 0$, and $\epsilon > 0$.
1. *Repeat*
      1. Determine z: Set $z = \mu m / \hat{\eta}$.

    2. Compute primal-dual search direction $\Delta \boldsymbol{y}_{\text{pd}}$ from (4-4).

    3. Line search and update: Determine step size $\gamma > 0$ and set $\boldsymbol{y} := \boldsymbol{y} + \gamma \Delta \boldsymbol{y}_{\text{pd}}$.

*Until* $\|r_{\text{pri}}\|_2 \leq \epsilon_{feas}$, $\|r_{\text{dual}}\|_2 \leq \epsilon_{feas}$, and $\hat{\eta} \leq \epsilon$.

2. *Return* $\boldsymbol{x}$, $\boldsymbol{\lambda}$, and $\boldsymbol{v}$.

In Step 1, the parameter $z$ is set to a factor $\mu$ times $m/\hat{\eta}$, which is the value of $z$ associated with the current surrogate duality gap $\hat{\eta}$. The algorithm terminates when $\boldsymbol{x}$ is primal feasible and $\boldsymbol{\lambda}, \boldsymbol{v}$ are dual feasible (within the tolerance $\epsilon_{feas}$), and the surrogate duality gap is smaller than the tolerance $\epsilon$. The following two subsections provide details required for implementation of the PD algorithm.

#### 4.1.3.1. Line search

The line search for the PD algorithm is a standard backtracking line search, based on the norm of the residual, and modified to ensure that $\boldsymbol{\lambda} > \boldsymbol{0}$ and $f(\boldsymbol{x}) < 0$. Let $\boldsymbol{x}, \boldsymbol{\lambda}$, and $\boldsymbol{v}$ denote the current iterate and $\boldsymbol{x}^+, \boldsymbol{\lambda}^+$, and $\boldsymbol{v}^+$ denote the next iterate, i.e.,

$$\boldsymbol{x}^+ = \boldsymbol{x} + \gamma \Delta \boldsymbol{x}_{\text{pd}}, \quad \boldsymbol{\lambda}^+ = \boldsymbol{\lambda} + \gamma \Delta \boldsymbol{\lambda}_{\text{pd}}, \quad \boldsymbol{v}^+ = \boldsymbol{v} + \gamma \Delta \boldsymbol{v}_{\text{pd}}.$$

One first computes the largest positive step size, not exceeding one, that gives $\boldsymbol{\lambda}^+ > \boldsymbol{0}$, i.e.,

$$\gamma^{\max} = \sup\{\gamma \in [0,1] | \boldsymbol{\lambda} + \gamma \Delta \boldsymbol{\lambda} \geq 0\}.$$

The backtracking procedure starts with $\gamma = 0.99 \gamma^{\max}$, and multiply $\gamma$ by $\beta \in (0,1)$ until $f(\boldsymbol{x}^+) < 0$. It continues multiplying $\gamma$ by $\beta$ until

$$\|r_z(\boldsymbol{x}^+, \boldsymbol{\lambda}^+, \boldsymbol{v}^+)\|_2 \leq (1 - \delta\gamma)\|r_z(\boldsymbol{x}, \boldsymbol{\lambda}, \boldsymbol{v})\|.$$

The backtracking parameters $\delta$ and $\beta$ are typically chosen in the ranges 0.01 to 0.1 and 0.3 to 0.8, respectively. It can be proved that the line search for the primal-dual method always terminates in a finite number of steps.

#### 4.1.3.2. Phase I

As mentioned above, the PD algorithm requires a strictly primal feasible starting point when finding a strictly primal feasible is not straightforward. When such a point is unknown, a preliminary stage, called the *phase I*, is carried out for the algorithm to compute this point (if there is one). The computed point is used as the starting point of the PD algorithm, which is called the *phase II*. For the phase I, a very common method is to form the problem

$$\begin{aligned} &\min \; s \\ &s.t. \\ &\quad f_i(\boldsymbol{x}) \leq s, \quad i = 1, \dots, m \\ &\quad \boldsymbol{Gx} = \boldsymbol{h} \end{aligned} \quad (4\text{-}6)$$

where $s$ is a new scalar variable. Assuming that a point $\boldsymbol{x}^0 \in \text{dom} f_1 \cap \dots, \text{dom} f_m$ with $\boldsymbol{Gx}^0 = \boldsymbol{h}$ is given, this problem is always strictly feasible with $s^0 > \max_{i=1,\dots,m} f_i(\boldsymbol{x}^0)$. Therefore, one can apply the PD algorithm to solve the above problem and find an optimal $\bar{x}$ and $\bar{s}$. If $\bar{s} < 0$, then $\bar{x}$ is a strictly feasible solution for the original problem (4-1). This phase can be skipped if finding an initial solution is straightforward.

**4.2. Proposed algorithm for portfolio optimization with EVaR**

Exploiting the particular structure of our optimization problem, this subsection shows how the PD algorithm can be efficiently implemented. To ease the presentation of our algorithm, without any loss of generality, we consider a linear portfolio with

$$G(\boldsymbol{w}, \boldsymbol{R}) = -\boldsymbol{R}^T \boldsymbol{w},$$

which is the most popular model in portfolio optimization literature. This specification enables readers to straightforwardly compare the design of the algorithms developed for EVaR-based and the CVaR-based portfolio optimization, considering that recent efficient algorithms for CVaR-based portfolio optimization are developed and tested only for linear portfolios.

Under the above setting, portfolio optimization with the EVaR becomes as follows:

$$\min t \ln \left( \sum_{j=1}^{N} p_j e^{-t^{-1}(a^j)^T \boldsymbol{w}} \right) - t \ln \alpha$$
$$\text{s.t.}$$
$$\boldsymbol{w}^T \boldsymbol{1} = 1$$
$$\boldsymbol{w} \in D$$
$$\boldsymbol{w} \geq \boldsymbol{0}, t > 0.$$
(4-7)

In the sequel, we give the details of our algorithm, called *EVaR-PD*, for solving this problem. The algorithm is designed based on the primal-dual method explained in Section 4.1. Let us define

$$\ln \boldsymbol{p} = \begin{bmatrix} \ln p_1 \\ \vdots \\ \ln p_N \end{bmatrix}, \text{ and } \boldsymbol{R}^S = \begin{bmatrix} \boldsymbol{a}^{1T} \\ \vdots \\ \boldsymbol{a}^{NT} \end{bmatrix}, \text{lse}(\boldsymbol{v}) = \ln(e^{v_1} + \cdots + e^{v_k}),$$

where $\boldsymbol{v}$ is a vector in $\mathbb{R}^k$. Then, the problem (3-7) can be compactly rewritten as

$$\min t \, \text{lse}\left( -\frac{\boldsymbol{R}^S \boldsymbol{w}}{t} + \ln \boldsymbol{p} \right) - t \ln \alpha$$
$$\text{s.t.}$$
$$\boldsymbol{w}^T \boldsymbol{1} = 1$$
$$\boldsymbol{w} \in D$$
$$\boldsymbol{w} \geq \boldsymbol{0}, t > 0.$$
(4-8)

Because $t > 0$ appears only in the objective function and affects no constraints, the feasibility problem associated with this problem does not involve variable $t$ and can be given by

$$\min s$$
$$\text{s.t.}$$
$$\boldsymbol{w}^T \boldsymbol{1} = 1$$
$$h_p(w) \le s, \quad p = 1, \dots, m \quad (4\text{-}9)$$
$$\boldsymbol{w} \ge \boldsymbol{0}.$$

Hence, an arbitrary positive value can be considered for $t$. Note that for $D = \mathbb{R}^n$, solving the feasibility problem associated with this problem is not required because a feasible solution can be easily determined.

We can now present the steps of the EVaR-PD algorithm for solving the problem (4-8), or equivalently, (4-7).

**EVaR-PD algorithm:**

1. *Input $n$, $\boldsymbol{R}^S$, $\boldsymbol{p}$, and $D$.*
2. Select $\boldsymbol{\lambda}^0 > \boldsymbol{0}$, $\boldsymbol{v}^0 > \boldsymbol{0}$, $t^0 > 0$, $\epsilon_{feas} > 0$, $\epsilon > 0$, and $\mu > 1$. Set $i = 1$, $t = t^0$, $\boldsymbol{\lambda} = \boldsymbol{\lambda}^0$ and $\boldsymbol{v} = \boldsymbol{v}^0$.
3. (Phase I) Select $t^0 > 0$ and $\boldsymbol{w}^0 \ge \boldsymbol{0}$ such that $\boldsymbol{w}^{0^T}\boldsymbol{1} = 1$. Let $\boldsymbol{x} = \begin{bmatrix} s^0 \\ \boldsymbol{w}^0 \end{bmatrix}$ with $s^0 > \max_{p=1,\dots,m} h_p(\boldsymbol{w}^0))$, and update $\boldsymbol{x}$ by applying the PD algorithm to solve the problem (4-9). Put $\boldsymbol{w}^0 = [x_2 \quad \cdots \quad x_{n+1}]^T$. For $D = \mathbb{R}^n$, skip solving the problem (4-9) and set $\boldsymbol{w}^0 = n^{-1}\boldsymbol{1}$.
4. (Phase II) Let $\boldsymbol{x} = \begin{bmatrix} \boldsymbol{w}^0 \\ t^0 \end{bmatrix}$ and update $\boldsymbol{x}$ by applying the PD algorithm to solve the problem (4-8), using the equations given in (4-10). Put $\boldsymbol{w} = [x_1 \quad \cdots \quad x_n]^T$.
5. *Output $\boldsymbol{w}$.*

One of the main computational difficulties for using the EVaR-PD algorithm in step 4 is forming the gradients and Hessians of the objective and constraint functions to compute the search direction. Fortunately, for our problem (4-8), these quantities can be given by simple closed-form expressions. For $D = \mathbb{R}^n$, these quantities are as follows:

$$f_0(\boldsymbol{w}, t) = t\,\text{lse}(\boldsymbol{y}) - t \ln \alpha$$

$$\nabla f_0(\boldsymbol{w}, t) = \begin{bmatrix} -\dfrac{(\boldsymbol{R}^S)^T e^{\boldsymbol{y}}}{\text{sum}(e^{\boldsymbol{y}})} \\ \text{lse}(\boldsymbol{y}) + \dfrac{\text{sum}(\boldsymbol{y}')}{\text{sum}(e^{\boldsymbol{y}})} - \ln \alpha \end{bmatrix}$$

$$\nabla^2 f_0(\boldsymbol{w}, t) = \begin{bmatrix} (\nabla^2 f_0)_1 & (\nabla^2 f_0)_2 \\ (\nabla^2 f_0)_3 & (\nabla^2 f_0)_4 \end{bmatrix} \quad (4\text{-}10)$$

$$(\nabla^2 f_0)_1 = \frac{1}{t}\left[\frac{(\boldsymbol{R}^S)^T \text{diag}(e^{\boldsymbol{y}}) \boldsymbol{R}^S}{\text{sum}(e^{\boldsymbol{y}})} - \frac{((e^{\boldsymbol{y}})^T \boldsymbol{R}^S)^T ((e^{\boldsymbol{y}})^T \boldsymbol{R}^S)}{\text{sum}(e^{\boldsymbol{y}})^2}\right]$$

$$(\nabla^2 f_0)_4 = \frac{1}{t}\left[\frac{\text{sum}(\boldsymbol{y}'')}{\text{sum}(e^{\boldsymbol{y}})} - \left(\frac{\text{sum}(\boldsymbol{y}')}{\text{sum}(e^{\boldsymbol{y}})}\right)^2\right]$$

$$(\nabla^2 f_0)_2^T = (\nabla^2 f_0)_3 = \left[ -\frac{y'^T R^S}{\text{sum}(e^y)} + \frac{(e^y)^T R^S \text{sum}(y')}{\text{sum}(e^y)^2} \right]$$

$$f(w, t) = \begin{bmatrix} -w \\ -t \end{bmatrix}$$

$$Df(w, t) = \begin{bmatrix} -\text{diag}(1) & 0 \\ 0 & -1 \end{bmatrix}, \quad i = 1, \ldots, n$$

$$\nabla^2 f_i(w, t) = 0, \quad i = 1, \ldots, n$$

where

$$y = -R^S \frac{w}{t} + \ln p, \ y' = \left(R^S \frac{w}{t}\right) \odot e^y = \text{diag}\left(R^S \frac{w}{t}\right) e^y, \ y'' = \left(R^S \frac{w}{t}\right) \odot y' = \text{diag}\left(R^S \frac{w}{t}\right) y'.$$

The symbol $\odot$ denotes the component-wise product, and $\text{sum}(y)$ and $\text{diag}(y)$ represent the sum of all entries of vector $y$ and a diagonal matrix having the entries of $y$ on its diagonal, respectively. Vectors $e^y$ and $\ln(y)$ denote the component-wise exponent and logarithm of $y$, i.e., $(e^y)_i = e^{y_i}$ and $(\ln y)_i = \ln y_i$, respectively. Each vector $e_i, i = 1, \ldots, n$, denotes an $n$-dimensional vector in which the $i$th element is one and the other elements are zero.

To see how the above computations changed by adding some new constraints, consider the case

$$D = \{w \in \mathbb{R}^n : \text{E}(-G(w, R)) = r^T w \geq r_{min}\}.$$

In this case, the constraint $r^T w \geq r_{min}$ where $r_{min}$ is a user-set minimum value for the expected portfolio return. Then, $f_0(w, t)$, $\nabla f_0(w, t)$, and $\nabla^2 f_0(w, t)$ are exactly formulated as in (4-10), and only the following minor modifications must be done:

$$f(w, t) = \begin{bmatrix} -w \\ -r^T w + r_{min} \\ -t \end{bmatrix}$$

$$Df(w, t) = \begin{bmatrix} -\text{diag}(1) & 0 \\ -r^T & 0 \\ 0 & -1 \end{bmatrix}, \quad i = 1, \ldots, n, \nabla f_i(w, t) = \begin{bmatrix} r \\ 0 \end{bmatrix}$$

$$\nabla^2 f_i(w, t) = 0, \quad i = 1, \ldots, n.$$

The high efficiency of the primal-dual method and the ease of computing the gradients and Hessians make the EVaR-PD algorithm very efficient for EVaR-based portfolio optimization in large scales. The computational efficiency of the EVaR-PD algorithm is evaluated in the next section.

## 5. Numerical study

In this section, we assess the efficiency of our proposed EVaR-PD algorithm for EVaR-based portfolio optimization, formulated in (4-7). Because this paper for the first time introduces a specialized algorithm for our problem, the computational performance of the EVaR-PD algorithm cannot be compared with an alternative specialized algorithm, and can be only compared by existing convex optimization software packages.

Our numerical study shows that the EVaR-PD algorithm <u>significantly</u> outperforms such general-purpose solvers. To carry out this comparison, the problem (4-7) was implemented using GAMS 24.1.2 API platform and solved using three nonlinear programming solvers: CONOPT, IPOPT, and PATHNLP. All the results are not reported here for the sake of brevity, but consider, for example, the <u>smallest</u> instance size given in Table 1.a with 50 instruments and 50000 samples. The average run times for solving 10 instances with this size using CONOPT, IPOPT, and PATHNLP solvers are 381, 199, and 221 seconds, respectively; while the mean time required by the EVaR-PD algorithm is only 11 seconds. This shows that our algorithm is <u>at least 19 times</u> faster than existing solvers for this instance size. For larger sizes, this superiority increases.

To provide a more strict evaluation of the EVaR-PD algorithm, its performance can be compared with that of efficient specialized algorithms developed for the CVaR-based portfolio optimization. In fact, both EVaR-based and CVaR-based approaches to portfolio optimization attempt to solve <u>the same financial decision problem</u> from different perspectives, where both take the same input data and return an investment portfolio (thought they solve different mathematical problems to find a portfolio). Therefore, one can compare their computational efficiency, which is an important factor for them to be used in practice by investment companies. Our numerical analysis reveals that EVaR-based portfolio optimization can be performed as fast as CVaR-based portfolio optimization in large scales. Moreover, as the sample size increases, the EVaR-based approach significantly performs better because its formulation size is independent of the sample size.

For linear portfolio optimization with the CVaR, recent studies show the fact that the LP method, resulting in the formulation (3-4), becomes inefficient as the sample size increases. Therefore, Lim et al. (2010), and Ogryczak and Śliwiński (2011) developed two efficient algorithms based on the different formulations (3-5) and (3-6), respectively. Their results showed that these algorithms were very effective and provided significant computational advantages over the LP method in large scales. Hence, in the following, these two algorithms are compared with the EVaR-PD algorithm. As stated in the introduction, there are many other papers proposing algorithms to solve CVaR-based portfolio optimization problems. However, based on the numerical results reported in these papers, it seems that these algorithms cannot considerably improve the computational efficiency of the algorithms proposed by Lim et al. (2010) and Ogryczak and Śliwiński (2011).

All algorithms were coded in C++ and CPLEX 9.0 C++ Concert Technology was used for solving the linear programs. All runs were implemented on a personal laptop with the Intel Core i5 2.53 GHz processor and Windows 7 with 4 GB of RAM. To have a fair comparison, the original code of the algorithm developed by Lim et al. (2010) was used, which was generously shared by the authors of this paper. Because for each instance size the test problems are randomly generated, the experiment was repeated 10 times, and then the average run times were reported. All times are reported in seconds.

**5.1. Algorithm setup**

An important setting in the EVaR-PD algorithm is the choice of the parameter $\mu$. From our experimentation, the value of $\mu = 5$ was found to work well for all instance sizes (the performance of the algorithm size-dependent values of $\mu$ may incrementally, not significantly, improve for some instances). The tolerance parameters $\epsilon_{feas}$ and $\epsilon$ in a primal-dual interior-point algorithm are usually chosen very small as convergence of this algorithm is often faster than linear. The values of these parameters are set to $10^{-6}$ for our algorithm. The parameter $\nu^0$ is chosen such that the surrogate duality gap becomes equal to 1, the parameter $\lambda^0$ is set to the unit vector, and the parameter $t^0$ is set to 1 as it was found to be suitable based on our experiments.

**5.2. Sets of test problems**

The EVaR-PD algorithm was fairly tested over test problems generated based on Lim et al. (2010). Note that test problems used in Ogryczak and Śliwiński (2011) were also taken from the paper Lim et al. (2010). In the following, we describe our test problems.

All random test instances used by Lim et al. (2010) are generated from a multivariate normal distribution. Considering that fat-tailed distributions are of interest in financial modeling, in addition to the normal distribution, we have also considered returns generated from the *t* distribution with degree of freedom (d.f.) $\nu = 5$, which is a fat-tailed distribution. It is notable that each instance with t distribution is obtained from its corresponding instance generated from the normal distribution.

To create a covariance matrix, Lim et al. (2010) generated a symmetric matrix whose off-diagonal elements were uniformly sampled from the interval [0,1], and the diagonal elements were then set so as to make the matrix diagonally dominant, which make the generated covariance matrices have a special structure. Hence, we generated the covariance matrices for five instances with the procedure used by Lim et al. (2010), and for the other five instances by multiplying a uniformly-generated random matrix by its transpose, which can generate arbitrary positive definite matrixes. In the following, the cases associated with these two covariance structures are distinguished by "Cov1" and "Cov2", respectively. When the results for the ten instances are reported together, the symbol "Cov" is used.

We generated three sets of random problem instances. The first set was used to study the run time trend as the number of instruments increases while sample size is fixed. In this set, the number of instruments was selected as 50, 100, 200, 750, and 1000, while the sample size was fixed to 50,000. The last two sets were used to study the trend as the sample size increases. For the second set, we fixed the number of instruments to 10 and generated instances having 100,000, 500,000, and 1,000,000 samples. In the third set, we fixed the number of instruments to 100 and generated instances having 50,000, 100,000, 150,000, 200000, and 750000 samples. Return rates in each case were generated from both normal and t distributions. Similarly as in Lim et al. (2010), the corresponding mean vector was set to zero, and the confidence level was selected as $1 - \alpha = 0.95$, for all instances.

**5.3. Comparison with algorithm proposed in Lim et al. (2010)**

Lim et al. (2010) studied NDO (Non-Differentiable Optimization) techniques that can be suitably employed to solve the non-differentiable formulation (3-6). They proposed a two-phase algorithm, which can be enhanced by another phase. This algorithm is called CVaR-IIP algorithm here. In the first phase, conventional differentiable optimization techniques are used while circumventing non-differentiable points. The second phase employs a variable target value NDO technique.

The optional third phase achieves an exact optimal solution in finite time by performing a switchover to a simplex solver starting with a crash basis obtained from the second phase. Their experimental results displayed that this third phase consumes a considerable amount of effort. On the other hand, the CVaR-IIP algorithm provides highly-accurate near-optimal solutions in most cases. Hence, they recommended employing the CVaR-IIP (and discarding the third phase) in practice as it can serve as an efficient solution procedure that provides a near-optimal solution in a timely manner for large-scale test problems.

Tables 1-3 present the computational results for examining the performance of the EVaR-PD algorithm versus the CVaR-IIP algorithm over the three sets of test problems described in Section 5.2. In these tables, each number under the column "GAP-IIP" represents the average of absolute optimality gap of the solutions obtained by the CVaR-IIP algorithm. For very large-sized instances, the optimality gaps for the CVaR-IIP algorithm were computed using a super computer with eight 3.2-GHz processors and 32 GB of RAM, based on the dual method given in the next subsection based on the formulation (3-5).

**Table 1.** Run times of algorithms EVaR-PD and CVaR-IIP, and the gap of the algorithm CVaR-IIP for the first set of test problems, given in Section 5-2, with fixed sample size $N = 50000$.

(**a**) Normal distribution

| $n$ | $N$ | EVaR-PD | | | CVaR-IIP | | | GAP-IIP | | |
|---|---|---|---|---|---|---|---|---|---|---|
| | | Cov1 | Cov2 | Cov | Cov1 | Cov2 | Cov | Cov1 | Cov2 | Cov |
| 50 | 50000 | 11 | 11 | 11 | 14 | 15 | 14 | 2.00E-06 | 0 | 1.00E-06 |
| 100 | 50000 | 25 | 26 | 25 | 28 | 28 | 28 | 1.00E-05 | 4.00E-06 | 7.00E-06 |
| 200 | 50000 | 69 | 77 | 73 | 59 | 58 | 59 | 0.000384 | 2.00E-05 | 0.0002 |
| 750 | 50000 | 1221 | 1286 | 1253 | 266 | 252 | 259 | 0.04142 | 0.00078 | 0.0211 |
| 1000 | 50000 | 3343 | 2976 | 3160 | 359 | 361 | 360 | 0.027742 | 0.00202 | 0.0149 |

(**b**) t distribution with d.f. $\nu = 5$

| $n$ | $N$ | EVaR-PD | | | CVaR-IIP | | | GAP-IIP | | |
|---|---|---|---|---|---|---|---|---|---|---|
| | | Cov1 | Cov2 | Cov | Cov1 | Cov2 | Cov | Cov1 | Cov2 | Cov |
| 50 | 50000 | 12 | 13 | 12 | 14 | 15 | 15 | 1.2E-05 | 0 | 6E-06 |
| 100 | 50000 | 26 | 30 | 28 | 28 | 29 | 28 | 0.000822 | 4E-05 | 0.0004 |
| 200 | 50000 | 66 | 79 | 72 | 58 | 58 | 58 | 0.035206 | 0.00014 | 0.0177 |
| 750 | 50000 | 849 | 1607 | 1228 | 266 | 276.92 | 271 | 0.448588 | 0.0126 | 0.2306 |
| 1000 | 50000 | 1672 | 5462 | 3567 | 382 | 405.7 | 394 | 0.528336 | 0.0247 | 0.2765 |

Figures 1-3 display the results related to the columns "Cov". From Figures 1, it seems that for the first set when the number of instruments increases, the CVaR-IIP algorithm requires less time than the EVaR-PD algorithm. However, one should carefully note that, the optimality gaps of the solutions obtained by this algorithm considerably increase (up to 0.528336), while for the EVaR-PD algorithm remain always less than $10^{-6}$.

**Table 2.** Run times of algorithms EVaR-PD and CVaR-IIP, and the optimality gaps of the algorithm CVaR-IIP for the second set of test problems, given in Section 5-2, with fixed number of instruments $n = 100$.

(**a**) Normal distribution

| $n$ | $N$ | EVaR-PD | | | CVaR-IIP | | | GAP-IIP | | |
|---|---|---|---|---|---|---|---|---|---|---|
| | | Cov1 | Cov2 | Cov | Cov1 | Cov2 | Cov | Cov1 | Cov2 | Cov |
| 100 | 50000 | 25 | 26 | 25 | 28 | 28 | 28 | 0.00049 | 4.2E-05 | 0.0003 |
| 100 | 100000 | 50 | 54 | 52 | 43 | 58 | 51 | 9.79E-05 | 0 | 5E-05 |
| 100 | 150000 | 85 | 96 | 91 | 102 | 90 | 96 | 0 | 0 | 0 |
| 100 | 200000 | 101 | 104 | 103 | 130 | 118 | 124 | 9.86E-05 | 0 | 5E-05 |
| 100 | 750000 | 401 | 433 | 417 | 491 | 480 | 486 | 2E-06 | 0 | 1E-06 |

(**b**) t distribution with d.f. $\nu = 5$

| $n$ | $N$ | EVaR-PD | | | CVaR-IIP | | | GAP-IIP | | |
|---|---|---|---|---|---|---|---|---|---|---|
| | | Cov1 | Cov2 | Cov | Cov1 | Cov2 | Cov | Cov1 | Cov2 | Cov |
| 100 | 50000 | 26 | 30 | 28 | 28 | 29 | 28 | 0.028883 | 0.0003 | 0.0146 |
| 100 | 100000 | 52 | 73 | 63 | 58 | 59 | 58 | 0.007702 | 0 | 0.0039 |
| 100 | 150000 | 75 | 117 | 96 | 91 | 102 | 96 | 0.002806 | 0 | 0.0014 |
| 100 | 200000 | 99 | 120 | 110 | 119 | 116 | 117 | 0.003154 | 0 | 0.0016 |
| 100 | 750000 | 416 | 520 | 468 | 480 | 484 | 482 | 0.000212 | 0 | 0.0001 |

From Figures 2 and 3, it is obvious that for the last two test sets the EVaR-PD algorithm outperforms the CVaR-IIP algorithm, and this relative advantage increases as the ratio of the sample size to the number of instruments increases. For these two sets, where the number of instruments is limited and fixed, the optimality gaps obtained by the CVaR-IIP algorithm are sufficiently good, but in most cases they are still larger than the EVaR-PD algorithm's accuracy level $10^{-6}$.

From Tables 2 and 3, one can also see that optimality gaps and run times slightly depend on the covariance structure. We can also see that there are no substantial differences in run times between the test problems with normal and t distribution of returns. However, from the first two sets it is evident that the optimality gaps for the CVaR-IIP algorithm significantly increase for the t distribution, which is heavy tailed.

**Table 3.** Run times for algorithms EVaR-PD and CVaR-IIP, and the optimality gaps of the algorithm CVaR-IIP for the third set of test problems, given in Section 5-1, with fixed number of instruments $n = 10$.

(**a**) Normal distribution

| $n$ | $N$ | EVaR-PD | | | CVaR-IIP | | | GAP-IIP | | |
|---|---|---|---|---|---|---|---|---|---|---|
| | | Cov1 | Cov2 | Cov | Cov1 | Cov2 | Cov | Cov1 | Cov2 | Cov |
| 10 | 100000 | 4 | 4 | 4 | 7 | 4 | 7 | 0 | 0 | 0 |
| 10 | 500000 | 20 | 20 | 20 | 32 | 6 | 32 | 0 | 0 | 0 |
| 10 | 1000000 | 42 | 43 | 43 | 66 | 32 | 65 | 0.07692 | 0 | 0.0385 |

(**b**) t distribution with d.f. $\nu = 5$

| $n$ | $N$ | EVaR-PD | | | CVaR-IIP | | | GAP-IIP | | |
|---|---|---|---|---|---|---|---|---|---|---|
| | | Cov1 | Cov2 | Cov | Cov1 | Cov2 | Cov | Cov1 | Cov2 | Cov |
| 10 | 100000 | 4 | 4 | 4 | 7 | 6 | 7 | 0 | 0 | 0 |
| 10 | 500000 | 20 | 21 | 20 | 32 | 31 | 32 | 0 | 0 | 0 |
| 10 | 1000000 | 40 | 43 | 42 | 66 | 62 | 64 | 0 | 0 | 0 |

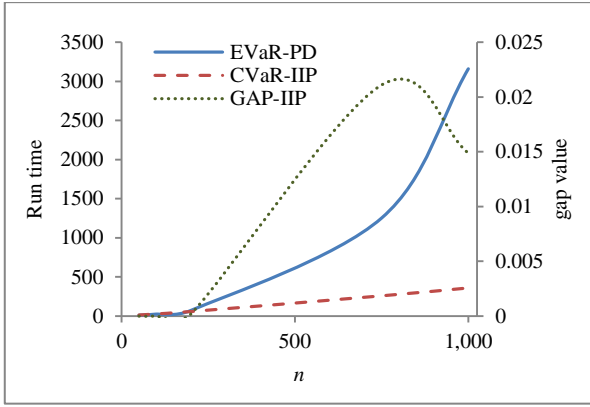 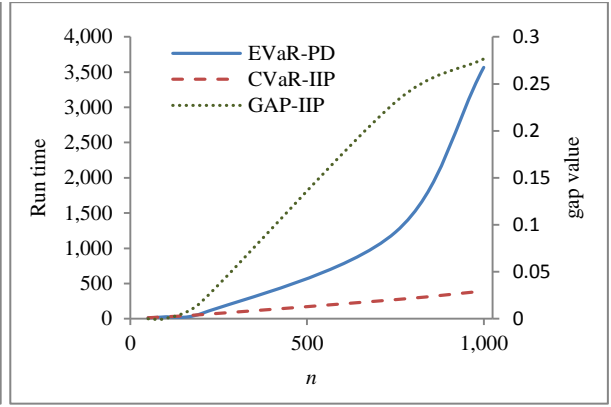

(**a**) Normal distribution  (**b**) t distribution with d.f. $\nu = 5$

**Figure 1.** Run times for algorithms EVaR-PD and CVaR-IIP, and the gap of the algorithm CVaR-IIP, versus the number of instruments $n$ for the first set of test problems, given in Section 5-2, with fixed sample size $N = 50000$.

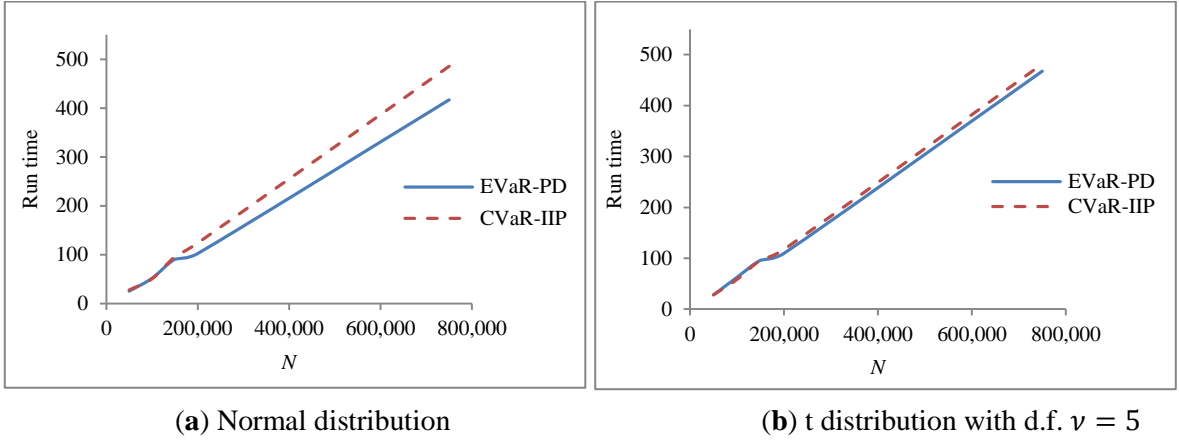

(**a**) Normal distribution        (**b**) t distribution with d.f. $\nu = 5$

**Figure 2.** Run times for algorithms EVaR-PD and CVaR-IIP versus the sample size for the second set of test problems, given in Section 5-2, with fixed number of instruments $n = 100$.

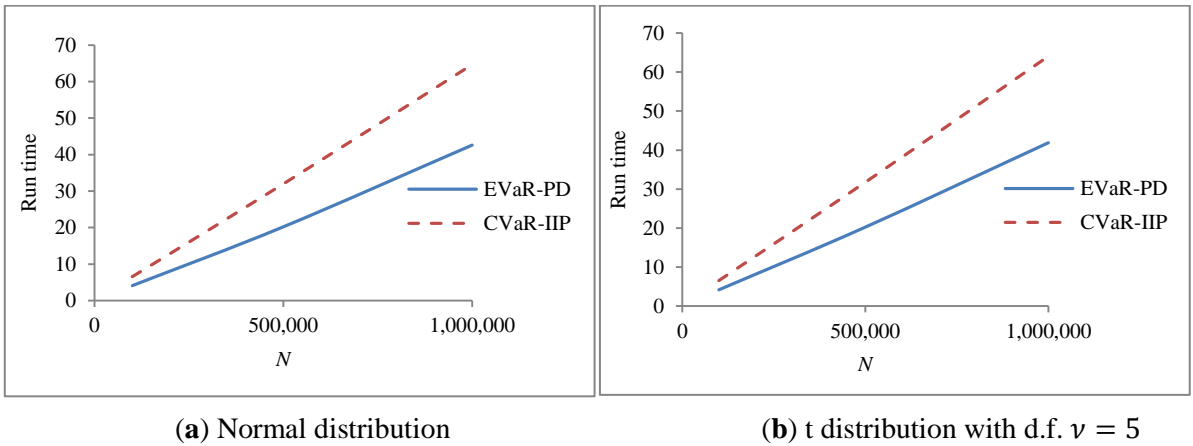

(**a**) Normal distribution        (**b**) t distribution with d.f. $\nu = 5$

**Figure 3.** Run times for algorithms EVaR-PD and CVaR-IIP versus the sample size for the third set of test problems, given in Section 5-2, with fixed number of instruments $n = 10$.

### 5.4. Comparison with algorithm of Ogryczak and Śliwiński (2011)

Ogryczak and Śliwiński (2011) showed that for solving the problem (3-4), the computational efficiency of the simplex method can be easily improved by using the LP duality, i.e., by solving the dual of the original problem (3-4), given in (3-5). Indeed, in the primal LP model (3-4), the number of constraints is proportional to the sample size, while in the dual program (3-5), the number of structural constraints is only proportional to the number of instruments. The resulting algorithm by the LP duality method is here called CVaR-DLP. The computational results of applying algorithms EVaR-PD and CVaR-DLP on the three sets of test problems described in Section 5.2 are given in Tables 4-6. Similar to the previous section, Figures 4-6 display the results related to the columns "Cov".

Note that the CVaR-DLP algorithm failed to solve the test problems with 750/ 50000, 1000/ 50000; and 100/ 750000 instruments/sample size; due to excessive memory requirements. These out-of-memory cases are indicated by "OM". As stated in the previous subsection, these cases were solved using a super computer within unlimited time to compute the optimality gaps of the algorithm CVaR-IIP.

From Tables 4-6, we can see that the CVaR-DLP algorithm requires more time than the EVaR-PD algorithm, and its required run time sharply increases as the ratio of the sample size to the number of instruments increases. Moreover, it can be seen that the CVaR-DLP algorithm preforms very differently for two kinds of covariance structures, which is not the case for the EVaR-PD algorithm.

**Table 4.** Run times for algorithms EVaR-PD and CVaR-DLP for the first set of test problems, given in Section 5-2, with fixed sample size $N = 50000$

(**a**) Normal distribution

| $n$ | $N$ | EVaR-PD | | | CVaR-DLP | | |
|---|---|---|---|---|---|---|---|
| | | Cov1 | Cov2 | Cov | Cov1 | Cov2 | Cov |
| 50 | 50000 | 11 | 11 | 11 | 31 | 11 | 21 |
| 100 | 50000 | 25 | 26 | 25 | 79 | 21 | 50 |
| 200 | 50000 | 69 | 77 | 73 | 277 | 47 | 162 |
| 750 | 50000 | 1221 | 1286 | 1253 | OM | OM | OM |
| 1000 | 50000 | 3343 | 2976 | 3160 | OM | OM | OM |

(**b**) t distribution with d.f. $v = 5$

| $n$ | $N$ | EVaR-PD | | | CVaR-DLP | | |
|---|---|---|---|---|---|---|---|
| | | Cov1 | Cov2 | Cov | Cov1 | Cov2 | Cov |
| 50 | 50000 | 12 | 13 | 12 | 32 | 11 | 22 |
| 100 | 50000 | 26 | 30 | 28 | 74 | 23 | 49 |
| 200 | 50000 | 66 | 79 | 72 | 271 | 46 | 159 |
| 750 | 50000 | 849 | 1607 | 1228 | OM | OM | OM |
| 1000 | 50000 | 1672 | 5462 | 3567 | OM | OM | OM |

**Table 5.** Run times for algorithms EVaR-PD and CVaR-DLP for the second set of test problems, given in Section 5-2, with fixed number of instruments $n = 100$.

(**a**) Normal distribution

| $n$ | $N$ | EVaR-PD | | | CVaR-DLP | | |
|---|---|---|---|---|---|---|---|
| | | Cov1 | Cov2 | Cov | Cov1 | Cov2 | Cov |
| 100 | 50000 | 25 | 26 | 25 | 79 | 21 | 50 |
| 100 | 100000 | 50 | 54 | 52 | 323 | 57 | 190 |
| 100 | 150000 | 85 | 96 | 91 | 655 | 99 | 377 |
| 100 | 200000 | 101 | 104 | 103 | 1073 | 129 | 601 |
| 100 | 750000 | 401 | 433 | 417 | OM | OM | OM |

(**b**) t distribution with d.f. $\nu = 5$

| $n$ | $N$ | EVaR-PD | | | CVaR-DLP | | |
|---|---|---|---|---|---|---|---|
| | | Cov1 | Cov2 | Cov | Cov1 | Cov2 | Cov |
| 100 | 50000 | 26 | 30 | 28 | 74 | 23 | 49 |
| 100 | 100000 | 52 | 73 | 63 | 302 | 50 | 176 |
| 100 | 150000 | 75 | 117 | 96 | 559 | 94 | 326 |
| 100 | 200000 | 99 | 120 | 110 | 785 | 121 | 453 |
| 100 | 750000 | 416 | 520 | 468 | OM | OM | OM |

**Table 6.** Run times for algorithms EVaR-PD and CVaR-DLP for the third set of test problems, given in Section 5-2, with fixed number of instruments $n = 10$.

(**a**) Normal distribution

| $n$ | $N$ | EVaR-PD | | | CVaR-DLP | | |
|---|---|---|---|---|---|---|---|
| | | Cov1 | Cov2 | Cov | Cov1 | Cov2 | Cov |
| 10 | 100000 | 4 | 4 | 4 | 23 | 7 | 15 |
| 10 | 500000 | 20 | 20 | 20 | 464 | 192 | 328 |
| 10 | 1000000 | 42 | 43 | 43 | 1522 | 592 | 1057 |

(**b**) t distribution with d.f. $\nu = 5$

| $n$ | $N$ | EVaR-PD | | | CVaR-DLP | | |
|---|---|---|---|---|---|---|---|
| | | Cov1 | Cov2 | Cov | Cov1 | Cov2 | Cov |
| 10 | 100000 | 4 | 4 | 4 | 30 | 13 | 21 |
| 10 | 500000 | 20 | 21 | 20 | 507 | 181 | 344 |
| 10 | 1000000 | 40 | 43 | 42 | 913 | 574 | 744 |

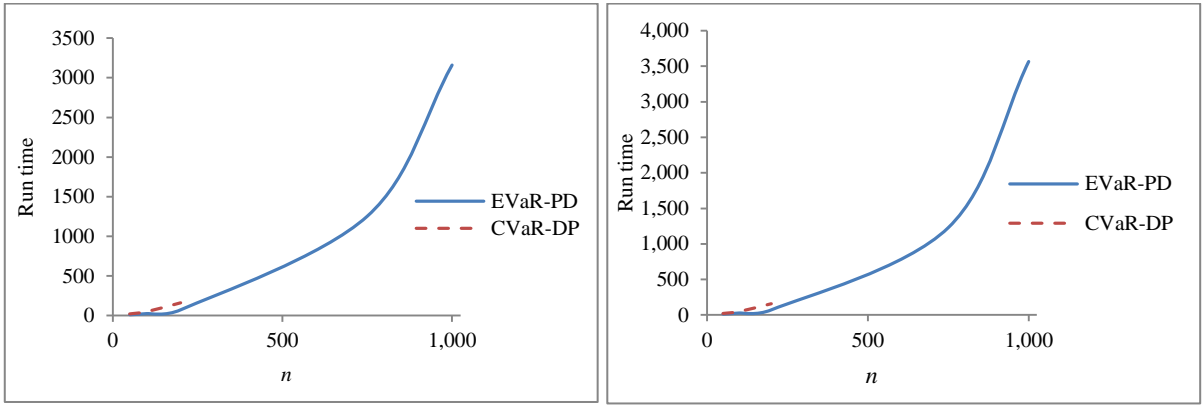

(**a**) Normal distribution  (**b**) t distribution with d.f. $\nu = 5$

**Figure 4.** Run times versus the number of instruments $n$ for algorithms EVaR-PD and CVaR-DLP for the first set of test problems, given in Section 5-2, with fixed sample size $N = 50000$.

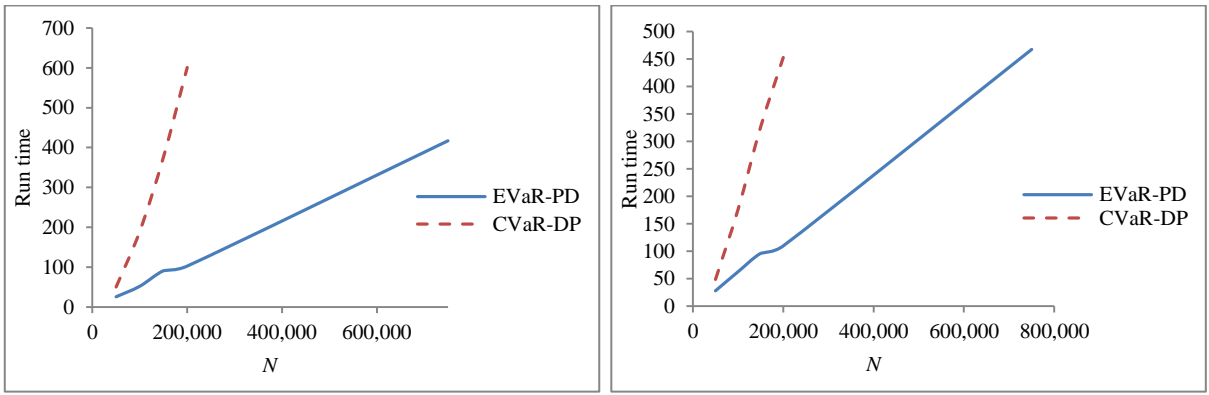

(**a**) Normal distribution  (**b**) t distribution with d.f. $\nu = 5$

**Figure 5.** Run times versus the sample size for algorithms EVaR-PD and CVaR-DLP for the second set of test problems, given in Section 5-2, with fixed number of instruments $n = 100$.

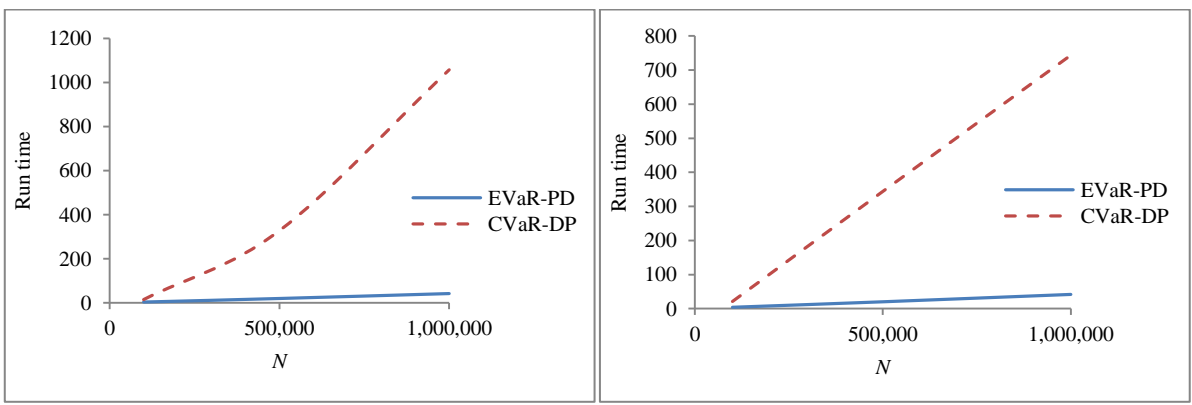

(**a**) Normal distribution  (**b**) t distribution with d.f. $\nu = 5$

**Figure 6.** Run times versus the sample size for algorithms EVaR-PD and CVaR-DLP for the third set of test problems, given in Section 5-2, with fixed number of instruments $n = 10$.

## 5.5. An investment case study

In this subsection, we compare the portfolios obtained by the EVaR and CVaR approaches on a real-world instance. The instance is constructed using data of S&P 500 (USA) capital market indices. We obtained daily opening price data from Yahoo Finance for the period of September 1984 to February 2015, which are available for 20 stocks of S&P 500 (i.e., stocks cat, aa, KO, DIS, GE, IBM, ed, cvx, aep, mo, cnp, UTX, MCD, GSPC, de, bmy, avp, axp, F, and WMT). Then, we calculated one-month return rates. This constructs an instance with $n = 20$ and $N = 15478$. The missing values were interpolated.

For this instance and different high confidence levels $1 - \alpha$, the optimal portfolios using the EVaR and CVaR approaches, denoted respectively by $\boldsymbol{w}_{EVaR,1-\alpha}$ and $\boldsymbol{w}_{CVaR,1-\alpha}$, were obtained. Table 7 compares these portfolios based on different traditional metrics: mean, standard deviation, and high-confidence VaR values. The number reported for each pair of confidence level and metric is the ratio of the value of that metric for the portfolio $\boldsymbol{w}_{EVaR,1-\alpha}$ to the value of that metric for the portfolio $\boldsymbol{w}_{CVaR,1-\alpha}$. For example, for row "0.95" and column "Mean", the number given in the corresponding cell is calculated as follows: $E(\boldsymbol{R}^T\boldsymbol{w}_{EVaR,0.95})/E(\boldsymbol{R}^T\boldsymbol{w}_{CVaR,0.95}) = 1.09$. The column "portfolio distance" reports the rectangular distance of $\boldsymbol{w}_{EVaR,1-\alpha}$ from $\boldsymbol{w}_{CVaR,1-\alpha}$, i.e., $\|\boldsymbol{w}_{EVaR,1-\alpha} - \boldsymbol{w}_{CVaR,1-\alpha}\|_1$.

From Table 7, first one can see that the portfolios obtained by the CVaR and EVaR approaches meaningfully differs because the numbers under the column "portfolio distance", ranging from 0.2 to 0.53, are significantly large compared to number 1, which represents the whole available capital. Secondly, for this instance the portfolios obtained by the EVaR approach have better means and VaR values in most cases, though the standard deviation values obtained for the EVaR portfolios are inferior. A reason for this may be that the EVaR is strongly and strictly monotone while the CVaR is not. These results suggest that the EVaR approach may result in more appropriate portfolios.

**Table 7.** Comparison of portfolios obtained by EVaR and CVaR approaches for a case study extracted from S&P 500.

| Confidence level | Mean | SD | VaR 0.99 | VaR 0.95 | VaR 0.90 | VaR 0.85 | VaR 0.80 | Portfolio distance |
|---|---|---|---|---|---|---|---|---|
| 0.99 | 1.04 | 1.01 | 1.01 | 1.00 | 0.99 | 0.99 | 1.02 | 0.20 |
| 0.95 | 1.09 | 1.04 | 1.08 | 1.02 | 0.99 | 1.01 | 1.04 | 0.34 |
| 0.90 | 1.15 | 1.05 | 1.08 | 1.04 | 1.03 | 1.03 | 1.04 | 0.28 |
| 0.85 | 1.40 | 1.15 | 1.20 | 1.09 | 1.06 | 1.09 | 1.13 | 0.53 |
| 0.80 | 1.29 | 1.05 | 1.08 | 1.06 | 1.05 | 1.08 | 1.06 | 0.33 |

## 6. Conclusion

This paper considers sample-based portfolio optimization with the EVaR. If the negative return is a differentiable convex function, the resulting optimization problem becomes a differentiable convex program where the number of constraints and variables is independent of the sample size. Therefore, portfolio optimization with the EVaR can be efficiently optimized using differentiable convex optimization techniques. This paper presents a primal-dual interior-point algorithm for portfolio optimization with the EVaR. A comprehensive computational study is conducted to test the proposed algorithm on three sets of test problems from the literature.

The proposed algorithm is much faster than the commercial solvers available for nonlinear optimization, and can stably solve problem instances with very large sizes ranging from 1000 instruments and 50,000 samples to 100 instruments and 750,000 samples, where the instances have different tail properties and dependence structures. Computational efficiency of CVaR-based portfolio optimization using two existing efficient algorithms is also compared with that of portfolio optimization with the EVaR using the proposed algorithm. This comparison shows that the EVaR approach to portfolio optimization can solve large-sized instances as fast as the CVaR approach does. Moreover, when the sample size increases, the EVaR approach outperforms the CVaR approach because the number of variables and constraints of the EVaR formulation is independent of the sample size.

The computational study on a real investment case indicates that the EVaR approach to portfolio optimization results in very different portfolios, compared to the ones obtained by the CVaR approach, which have better traditional performance metrics, such as mean and high-confidence VaR values.

Given that the EVaR is coherent and has more appropriate monotonicity properties compared to the VaR and CVaR as well as its computational superiorities, the EVaR can be a promising risk measure from both financial and computational perspectives.

Although the proposed algorithm is efficient in large scales, developing more efficient algorithms is a future research direction. Optimization of other portfolio optimization problems with the EVaR could also be investigated in future studies.

**Acknowledgements**

The authors would like to thoroughly thank the authors of the paper Lim et al. (2010), especially Dr. Churlzu Lim, for sharing the code of the algorithm proposed in their paper and for their helpful supports.